\documentclass[aps,prl,twocolumn,superscriptaddress,groupedaddress]{revtex4}  
\usepackage{times}
\usepackage{graphicx}  
\usepackage{graphics}
\usepackage{dcolumn}   
\usepackage{bm}        
\usepackage{amssymb}   
\usepackage{siunitx}   
\usepackage{amsmath}
\usepackage{amssymb}
\usepackage{float}
\usepackage[colorlinks,linkcolor=blue,citecolor=blue]{hyperref}

\begin{document}

\widetext

\title{\textbf{On-demand Semiconductor Source of Entangled Photons Which Simultaneously Has \\ High Fidelity, Efficiency, and Indistinguishability}}

\author{Hui Wang}
\affiliation{%
Shanghai Branch, National Laboratory for Physical Sciences at Microscale,
    	University of Science and Technology of China, Shanghai 201315, China%
	}
\affiliation{%
CAS Center for Excellence and Synergetic Innovation Center in Quantum Information and Quantum Physics, University of Science and Technology of China, Hefei, Anhui 230026, China%
    }

\author{Hai Hu}
\affiliation{%
Division of Nanophotonics, CAS Center for Excellence in Nanoscience, National Center for Nanoscience and Technology, Beijing, 100190, China%
    }

\author{T.-H. Chung}
\author{Jian Qin}
\affiliation{%
Shanghai Branch, National Laboratory for Physical Sciences at Microscale,
    	University of Science and Technology of China, Shanghai 201315, China%
	}
\affiliation{%
CAS Center for Excellence and Synergetic Innovation Center in Quantum Information and Quantum Physics, University of Science and Technology of China, Hefei, Anhui 230026, China%
    }

\author{Xiaoxia Yang}
\affiliation{%
Division of Nanophotonics, CAS Center for Excellence in Nanoscience, National Center for Nanoscience and Technology, Beijing, 100190, China%
    }
\author{J.-P. Li}
\author{R.-Z. Liu}
\author{H.-S. Zhong}
\author{Y.-M. He}
\author{Xing Ding}
\author{Y.-H. Deng}
\affiliation{%
Shanghai Branch, National Laboratory for Physical Sciences at Microscale,
    	University of Science and Technology of China, Shanghai 201315, China%
	}
\affiliation{%
CAS Center for Excellence and Synergetic Innovation Center in Quantum Information and Quantum Physics, University of Science and Technology of China, Hefei, Anhui 230026, China%
    }

\author{C. Schneider}
\affiliation{%
Technische Physik, Physikalisches Instit{\"a}t and Wilhelm Conrad R{\"o}ntgen-Center for Complex Material Systems, Universitat W{\"u}rzburg, Am Hubland, D-97074 W{\"u}zburg, Germany%
	}

\author{Qing Dai}\email{daiq@nanoctr.cn}
\affiliation{%
Division of Nanophotonics, CAS Center for Excellence in Nanoscience, National Center for Nanoscience and Technology, Beijing, 100190, China%
    }

\author{Y.-H. Huo}\email{yongheng@ustc.edu.cn}
\affiliation{%
Shanghai Branch, National Laboratory for Physical Sciences at Microscale,
    	University of Science and Technology of China, Shanghai 201315, China%
	}
\affiliation{%
CAS Center for Excellence and Synergetic Innovation Center in Quantum Information and Quantum Physics, University of Science and Technology of China, Hefei, Anhui 230026, China%
    }

\author{Sven H{\"o}fling}
\affiliation{%
Shanghai Branch, National Laboratory for Physical Sciences at Microscale,
    	University of Science and Technology of China, Shanghai 201315, China%
    }
\affiliation{%
Technische Physik, Physikalisches Instit{\"a}t and Wilhelm Conrad R{\"o}ntgen-Center for Complex Material Systems, Universitat W{\"u}rzburg, Am Hubland, D-97074 W{\"u}zburg, Germany%
    }
\affiliation{%
SUPA, School of Physics and Astronomy, University of St Andrews, St Andrews KY16 9SS, United Kingdom%
    }
\author{Chao-Yang Lu}\email{cylu@ustc.edu.cn}
\author{Jian-Wei Pan}\email{pan@ustc.edu.cn}
\affiliation{%
Shanghai Branch, National Laboratory for Physical Sciences at Microscale,
    	University of Science and Technology of China, Shanghai 201315, China%
	}
\affiliation{%
CAS Center for Excellence and Synergetic Innovation Center in Quantum Information and Quantum Physics, University of Science and Technology of China, Hefei, Anhui 230026, China%
    }
\date{\today}

\begin{abstract}
An outstanding goal in quantum optics and scalable photonic quantum technology is to develop a source that each time emits one and only one entangled photon pair with simultaneously high entanglement fidelity, extraction efficiency, and photon indistinguishability. By coherent two-photon excitation of a single InGaAs quantum dot coupled to a circular Bragg grating bullseye cavity with broadband high Purcell factor up to 11.3, we generate entangled photon pairs with a state fidelity of 0.90(1), pair generation rate of 0.59(1), pair extraction efficiency of 0.62(6), and photon indistinguishability of 0.90(1) simultaneously. Our work will open up many applications in high-efficiency multi-photon experiments and solid-state quantum repeaters.
\end{abstract}

\maketitle

Quantum entanglement \cite{Schrodinger1935} between flying photons \cite{wu1950} are central in the Bell test \cite{Bell1964} of the contradiction between local hidden variable theory and quantum mechanics \cite{EPR1935}. Aside from the fundamental interest, the entangled photons have been recognized as the elementary resources in quantum key distribution \cite{Ekert1991}, quantum teleportation \cite{Bennett1993}, quantum metrology \cite{Metrology2006} and quantum computing \cite{Raussendorf2001}. There has been a strong interest in experimental generations of entangled photons from trapped atoms \cite{Freedman1972}, spontaneous parametric down-conversion (SPDC) \cite{Kwiat1995}, and quantum dots \cite{Benson2000} etc. A checklist of relevant parameters for an entangled photon source include \cite{Lu2014}:

  {\bf A}. Entanglement fidelity. The produced two photons should be in a state close to a maximally entangled Bell state.

  {\bf B}. On-demand generation. The source should, at a certain time, emit one and only one pair of entangled photons.

  {\bf C}. Extraction efficiency. The photons should be extracted out from the source and collected with a high efficiency.

  {\bf D}. Indistinguishability. The photons emitted from different trials should be exactly identical in all degrees of freedom.

The past decades witnessed increasingly more sophisticated Bell tests and advanced photonic quantum information technologies enabled by developments of the photon entanglement source striving to fulfill the four checklist. For example, by combining {\bf A} and {\bf C}, the SPDC photons allowed for Bell tests closing both the locality and detection loopholes simultaneously \cite{Giustina2015,Shalm2015}. Very recently, ultrafast pulsed SPDC satisfied {\bf A}, {\bf C}, and {\bf D} and was exploited to demonstrate 12-photon entanglement and scattershot boson sampling \cite{Zhong2018}. However, the item {\bf B} remains an intrinsic problem for the SPDC where the photon pairs are generated probabilistically, and inevitably accompanied with undesirable multi-pair emissions.

An alternative route to generate entangled photons is through radiative cascades in single quantum emitters such as quantum dots which can have a near-unity quantum efficiency \cite{Benson2000}, therefore meet the item {\bf B}. However, the solid-state artificial atom system has its own challenges, including the structural symmetry, extraction efficiency, and dephasings. To this end, tremendous progress has been reported in eliminating the fine structure splitting of neutral excitons \cite{Akppian2006,Huo2013,Huber2018}, improving the extraction efficiency using double-micropillar structures \cite{Dousse2010} or broadband antennas \cite{Versteegh2014,Jons2017,Chen2018}, and enhancing the entanglement fidelity and photon indistinguishability using resonant excitation \cite{Muller2014,Stevenson2012}. Encouragingly, the entanglement fidelity ({\bf A}) and the photon indistinguishability ({\bf D}) (for \SI{2}{ns} separation) has reached 0.978(5) and 0.93(7), respectively \cite{Huber2018,Stevenson2012}. Very recently, the entanglement fidelity of 0.9 ({\bf A}) was combined with a record-high pair efficiency of 0.37 per pulse, which is the product of the pair generation rate of 0.88 (\textbf{B}) and the extraction efficiency of 0.42 (\textbf{C}), on the same device \cite{Chen2018}.

In this Letter, we report a near-perfect entangled-photon source that for the first time fulfills {\bf A}-{\bf D}. By coherently driving a single InGaAs quantum dot coupled to a bullseye microcavity with broadband Purcell enhancement, we create entangled photons with a fidelity of 0.90(1), pair generation rate of 0.59(1), pair extraction efficiency of 0.62(6), and photon indistinguishability of 0.90(1) simultaneously.

While polarized single-photon sources from quantum dot-micropillars with both high efficiency and photon indistinguishability have been demonstrated very recently \cite{Ding2016}, the creation of near-perfect entangled photon pairs posed additional challenges. First, the fine structure splitting should be smaller than the radiative linewidth of the single photons, leaving no room for leaking which-path information. Second, as the two single photons from the biexciton-exciton (XX-X) radiative cascaded emission have different wavelengths, broadband-Purcell cavities should be used to enhance both the XX and X photons. The Purcell factor that can accelerate the radiative decay rate, together with resonant excitation without inducing dephasing and emission time jitter, is desirable both for improving the two-photon entanglement fidelity and indistinguishability.

\begin{figure}
  \centering
  \includegraphics[width=0.45\textwidth]{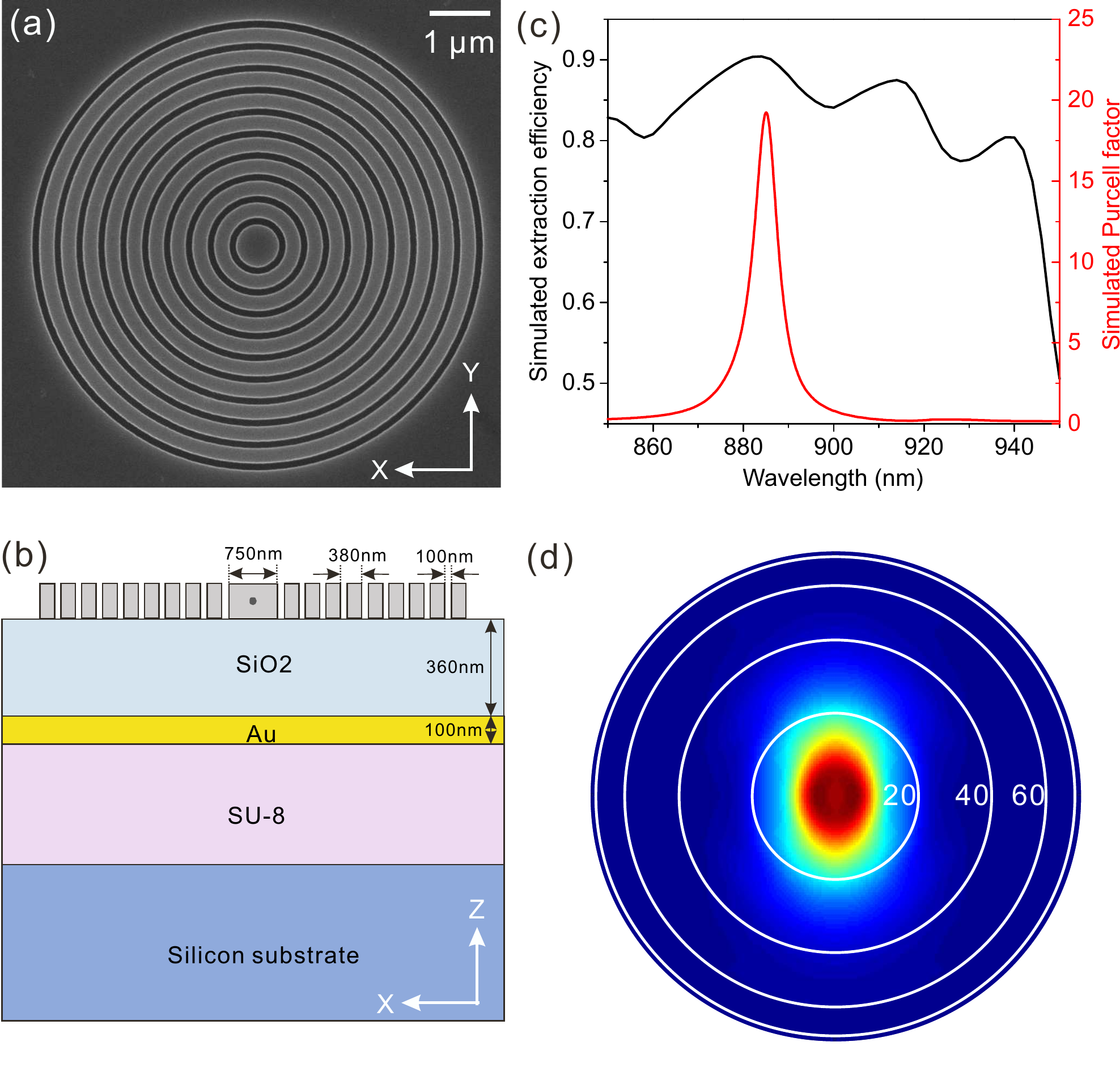}
  \caption{Nanostructure and simulation of circular Bragg grating (CBG) cavity in a bullseye geometry. (a) Top view of a scanning electron microscope image of the CBG cavity in X-Y plane. (b) Side view of our device. The design parameters of the CBG are labelled. (c) Numerical simulation of the single-photon extraction efficiency and Purcell factor as a function of photon emission wavelength indicates a broadband feature of the CBG cavity. (d) Numerical simulation of far-field distribution of the electrical intensity of the emission assuming an emitter sitting in the center of the CBG.}
  \label{Fig:1}
\end{figure}

\begin{figure*}
  \centering
  \includegraphics[width=0.65\textwidth]{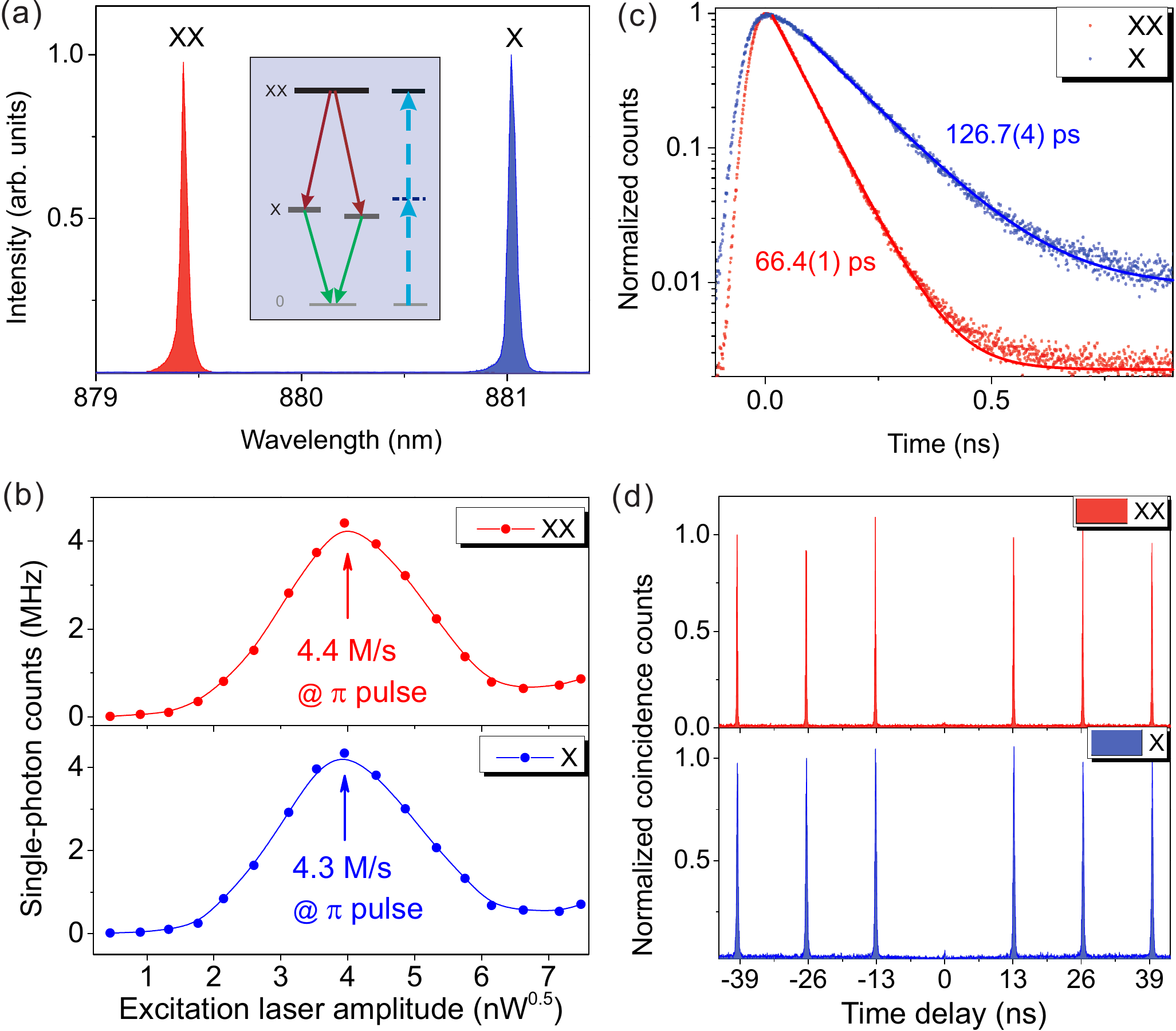}
  \caption{Brightness and purity of the XX and X photons from an InGaAs quantum dot in CBG. (a) The spectrum of the cascaded emitted XX and X photons from the level structure shown in the inset. The energy of the pulsed laser for excitation is set at the average energy of the XX and X photons, in resonance with the virtual biexciton two-photon excitation state. (b) The eventually detected single photon counts as a function of the square root of excitation laser power, showing a clear Rabi oscillation. (c) Measurement of time-resolved XX and X photon counts to determine their lifetime. (d) Intensity-correlation histogram of the XX and X photons under ¦Ð pulse excitation, obtained using a Hanbury Brown and Twiss-type setup.}
  \label{Fig:2}
\end{figure*}

We choose self-assembled InGaAs quantum dots as single quantum emitters which have near-unity quantum efficiencies \cite{Shields2007}---a prerequisite for the criteria {\bf B}---and near-transform-limited emission linewidth \cite{Kuhlmann2015}. For a broadband high-Purcell cavity, we adopt circular Bragg grating (CBG) in a bullseye geometry \cite{Green2004} which features a small effective mode volume and a relatively low Q factor ($\sim$150). The CBGs have been previously employed to enhance the single-photon collection from quantum dots \cite{Davanco2011} and nitrogen vacancy centers in diamond \cite{Li2015}. A scanning electron microscope image of our CBG device is shown in Fig.~\ref{Fig:1}a. We design the parameters of the CBG as detailed in Fig.~\ref{Fig:1}b in order to align its resonance with a moderate spectral range of $\sim$\SI{5}{nm} to the center of the wavelength of the photon pairs (see the caption of Fig.~\ref{Fig:1} and \cite{Supplement}).

To redirect the single photon emission from downward back to upward, a gold mirror is fabricated at the bottom of the quantum dot. The idea of backside metallic broadband mirror has been proposed in solid immersion lens and antennas \cite{Chen2018}, nanowires \cite{Reimer2012}, and others \cite{Chen2011}. A \SI{360}{nm} thick ${\rm SiO_{2}}$ is sandwiched between the GaAs membrane and the gold mirror, forming a constructive interference between the downward and upward light. Our numerical simulation in Fig.~\ref{Fig:1}c shows that for our chosen parameters, a Purcell factor of $\sim$20 and an extraction efficiency (defined as the ratio of single photons that had been generated by the quantum dot to those escaped from the bulk GaAs and were collection into the first lens) up to 90$\%$ can be achieved for both the X and XX photons. Another key issue to check is that whether the emitted photons can be efficiently collected into a single-mode fiber. We simulate the far field intensity distribution using finite-different time-domain method. The numerical results (see Fig.~\ref{Fig:1}d) shows that the single-photon emission is highly directional and slightly elliptical. An objective lens with a numerical aperture (NA) of 0.65 is capable of collecting $\sim$90$\%$ of the emitted photons.

\begin{figure*}
  \centering
  \includegraphics[width=0.86\textwidth]{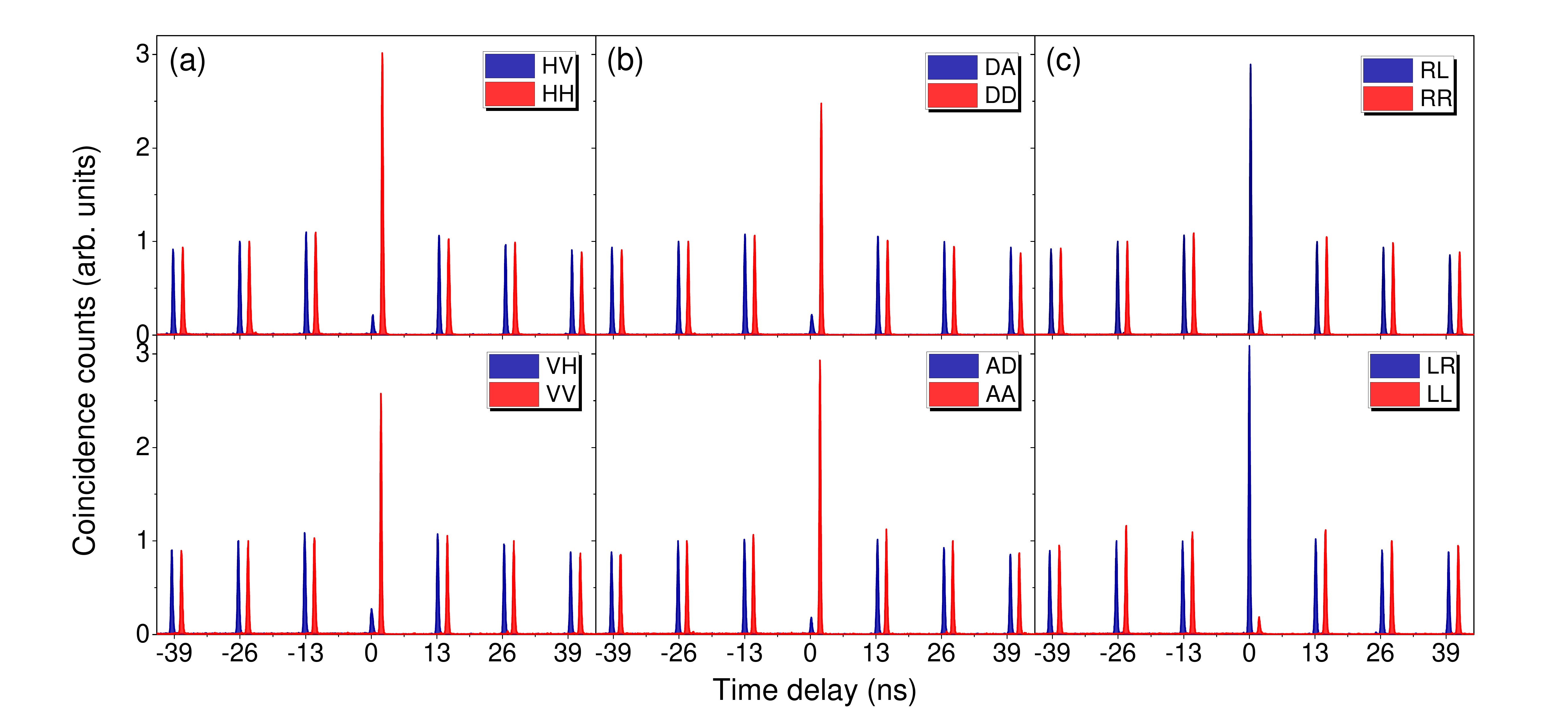}
  \caption{Measurement of two-photon entanglement fidelity. (a) Detected XX-X cross-correlation coincidence counts in linear basis. H: horizontal. V: vertical. (b) in diagonal basis. D: +45¡ã. A: -45¡ã. (c) in circular basis. R: right circular. L: left circular. All the photon counts within 2 ns time bin are used for the calculation of the entanglement fidelity. Note that the ratio of the sum of the two peaks at $\sim$\SI{13}{ns} to the sum of the two peaks at zero time can be used to reliably extract the XX excited-state preparation efficiency and quantum radiative efficiency (see \cite{Supplement} for details).}
  \label{Fig:3}
\end{figure*}

As illustrated in the inset of Fig.~\ref{Fig:2}a, our scheme to generate entangled photons is via XX-X cascade radiative decay in an InGaAs quantum dot. The polarization of emitted photons is determined by the spin of the intermediate exciton states. In our sample, $\sim$3$\%$ of the quantum dots show fine structure splitting below \SI{2.5}{\mu eV}. We pick a quantum dot with a small fine structure splitting of $\textless$\SI{1.2}{\mu eV}, which is limited by the resolution of the spectrometer. We use coherent two-photon excitation scheme \cite{Muller2014} to pump the quantum dot to the XX state. The energy of the pump pulsed laser is set at the average energy of the XX and X photons. We observe a clean photon pair emission spectrum as shown in Fig.~\ref{Fig:2}a, where the X and XX lines are separated by $\sim$\SI{1.6}{nm}.

\begin{figure*}
  \centering
  \includegraphics[width=0.80\textwidth]{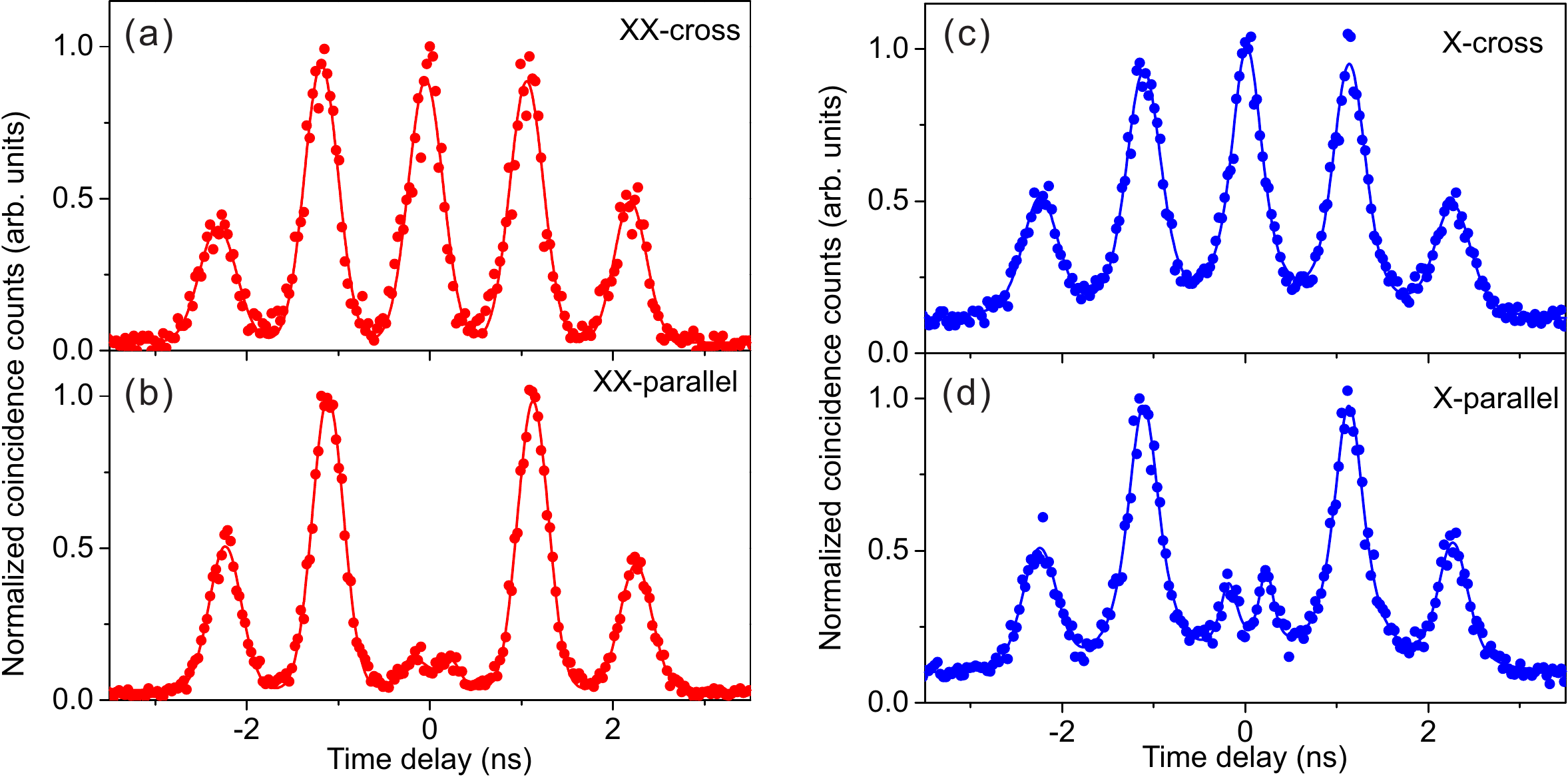}
  \caption{The interference between two XX photon is plotted in (a) and (b). The same data for X photons is shown in (c) and (d). The input two photons are ¦Ð-pulse excited and prepared in cross (a), (c) and parallel (b), (d) polarizations, respectively. The fitting function is the convolution of exponential decay (emitter decay response) with Gaussian (photon detection time response). All the data points presented are raw data without background subtraction.}
  \label{Fig:4}
\end{figure*}

We vary the average power of the laser and record the photon counts with a superconducting nanowie single-photon detector. The data for both XX and X photons are shown in Fig.~\ref{Fig:2}b, where we observe clear Rabi oscillations due to a coherent control of the quantum dot biexcitonic system \cite{Muller2014}. The XX and X photon count rates reach their first maxima at $\pi$ pulses under a pumping laser power of $\sim$\SI{16}{nW}. Such a power is about 3 orders of magnitudes lower than those in non-resonant excitation where the photon counts usually grow asymptotically with pump power \cite{Akppian2006,Dousse2010}. The efficient excitation requiring only very low power is important for eliminating the undesired multiexciton states and fluctuating electrical noise in the vicinity of the quantum dot.

Under a pumping rate of \SI{76}{MHz} and at $\pi$ pulse, the final count rates observed in our experimental set-up are $4.41\times10^{6}/\mathrm{s}$ and $4.34\times10^{6}/\mathrm{s}$ for the XX and X photons, respectively. The two-photon coincidence rate is $4.20\times10^{5}/\mathrm{s}$. The Klyschko efficiency for the XX (X) photons are $9.5\%$ ($9.7\%$). We discuss in \cite{Supplement} on current limitations and possible improvements to increase the two-photon coincidence rate. By bookkeeping independently calibrated single-photon detection efficiency ($\sim$76$\%$), optical path transmission rate ($\sim$25$\%$, including optical window, grating, two beam splitters and fiber connectors), and single-mode fiber coupling efficiency ($\sim$65$\%$), XX excited-state preparation efficiency at $\pi$ pulse and radiation efficiency ($\sim$70$\%$), blinking (occurring with a probability of 16$\%$), we estimate that $\sim$79.5$\%$ ($\sim$78.2$\%$) of the generated XX (X) single photons are collected into the first objective lens (NA=0.68)  \cite{Supplement}. This corresponds to a record high photon pair extraction efficiency (which is the product of the two single-photon extraction efficiencies) of 0.62(6) (criteria {\bf C}).

The record high photon counts observed in Fig.~\ref{Fig:2}b suggest a strong Purcell coupling with single quantum emitters. To quantify the Purcell factor, we perform time-resolved resonance fluorescence measurements under the two-photon excitation to extract the radiative lifetimes of the XX and X photons (Fig.~\ref{Fig:2}c), which are \SI{66.4(1)}{ps} and \SI{126.7(4)}{ps}, respectively, shortened by a factor of 11.3 and 8.7 compared to the quantum dot in bulk GaAs. The Purcell factor of the XX photon is higher than that of the X photon, which is due to a better spectral match to the cavity \cite{Supplement}. The CBG cavity not only gives comparable Purcell factors as those in the state-of-the-art micropillar-quantum dot single-photon devices \cite{Ding2016}, more interestingly, it also works over a moderately broadband over a few nanometers.

The XX and X photons are first characterized separately by second-order correlation measurements. Owing to the two-photon excitation scheme that spectrally separates the scattering laser from the emitted photons, near background-free entangled photons can be obtained \cite{Schweickert2018}. This is confirmed by the accumulated intensity-correlation histogram in Fig.~\ref{Fig:2}d, where at $\pi$ pulse, nearly vanishing double-photon emission probabilities,  $g^{(2)}_{XX}(0)=0.014(1)$, and $g^{(2)}_{X}(0)=0.013(1)$, are observed at zero time delay without any background subtraction. The strong anti-bunching reveals a near-perfect single-photon nature even under saturation pumping, without any fundamental tradeoff between the generation efficiency and the single-photon purity (criteria {\bf B}), an intrinsic advantage compared to the parametric down-conversion \cite{Kwiat1995} where increasing the photon pair rate inevitably induces more higher-order photon emission.

Next, we characterize the entangled photons by measuring their state fidelity, that is, the overlap of our experimentally produced states with an ideal, maximally entangled state (criteria {\bf A}). We perform polarization-resolved cross-correlation measurements between the XX and X photons. The correlations at three mutually unbiased basis, right (R) and left (L) circular, horizontal (H) and vertical (V), diagonal (D) and anti-diagonal (A), and are plotted in Fig.~\ref{Fig:3}. In the linear and diagonal basis (see Fig.~\ref{Fig:3}a and \ref{Fig:3}b), the measured histograms show a strong bunching when XX and X photons have parallel polarizations and an antibunching when they are orthogonal. The data in the circular basis (see Fig.~\ref{Fig:3}c) shows an opposite behavior. The data suggests that the entangled two-photon state is close to the form of $|\psi\rangle_{XX,X}=(|H\rangle_{XX}|H\rangle_{X}+|V\rangle_{XX}|V\rangle_{X})/\sqrt{2}$. The correlation visibilities for each basis should be calculated by:
\begin{equation}
V_{\mathrm{basis}}=\frac{g^{(2)}_{XX,X}(0)-g^{(2)}_{XX,\overline{X}}(0)-g^{(2)}_{\overline{XX},X}(0)+g^{(2)}_{\overline{XX},\overline{X}}(0)}{g^{(2)}_{XX,X}(0)+g^{(2)}_{XX,\overline{X}}(0)+g^{(2)}_{\overline{XX},X}(0)+g^{(2)}_{\overline{XX},\overline{X}}(0)} \nonumber
\end{equation}
where $g^{(2)}_{XX,X}(0)$, $g^{(2)}_{\overline{XX},\overline{X}}(0)$ are the coincidences of co-polarized bases, and $g^{(2)}_{XX,\overline{X}}(0)$, $g^{(2)}_{\overline{XX},X}(0)$ are those of cross-polarized bases. From a complete and necessary set of 12 measurements (note that the 12 measurements are both sufficient and necessary to reliably derive the two-photon state fidelity, without additional assumptions of H/V, D/A, R/L symmetry) as plotted in Fig.~\ref{Fig:3}, we extract $V_{\mathrm{linear}}=0.84(1)$, $V_{\mathrm{diagonal}}=0.86(1)$, and $V_{\mathrm{circular}}=-0.88(1)$. Thus, the fidelity to the maximally entangled state is obtained as $F=(1+V_{\mathrm{linear}}+V_{\mathrm{diagonal}}-V_{\mathrm{circular}})/4=0.90(1)$. We note that here the high Purcell factor broadens the intrinsic linewidth of the photons and thus a larger fine structure splitting can be tolerated, which is favorable for a high-fidelity two-photon entanglement. The residual fine structure splitting can be further eliminated to nearly zero by strain tuning \cite{Huber2018}, a technique perfectly compatible with the current membrane structure.

Having simultaneously fulfilled the criteria {\bf A}, {\bf B}, and {\bf C}, finally we turn to test the {\bf D}: photon indistinguishability. For this measurement, the quantum dot is excited every \SI{13.1}{ns} by two $\pi$ pulses separated by \SI{1}{ ns}. Hong-Ou-Mandel interference between the two consecutive photons is performed using an unbalanced Mach-Zehnder interferometry setup, as in Ref. \cite{Ding2016}, in parallel and orthogonal polarization configurations. The outputs of this interferometer are detected by single-mode fiber-coupled single-photon counters. For both XX and X photon, a record of coincidence events is kept to build up a time-delayed histogram as shown in Fig.~\ref{Fig:4}. In both cases, we observe a strong suppression of the coincidences at zero delay when the two photons are in the parallel polarization.

Raw interference visibilities calculated from the areas of the central peaks by
 \begin{equation}
 V_{\mathrm{HOM}}=\frac{g^{(2)}_{\mathrm{cross}}(0)-g^{(2)}_{\mathrm{parallel}}(0)} {g^{(2)}_{\mathrm{cross}}(0)} \nonumber
\end{equation}
for the XX and X photons are 0.86(1) and 0.67(1), respectively. Taking into account of the residual two-photon events probability and the independently calibrated optical imperfections of our interferometric set-up, a corrected degree of indistinguishability for the XX and X photons are estimated to be 0.90(1) and 0.71(1), respectively. The data for pulse separation of 2 ns and 13 ns are presented in \cite{Supplement}. A closer inspection of the coincidence counts in Fig.~\ref{Fig:4}b-d shows a dip around the zero delay due to temporal filtering by ultrafast timing resolution ($\sim$\SI{20}{ ps}) of the superconducting nanowire single-photon detectors, which improve the interference visibility to 0.93(2) and 0.86(3) for the XX and X photons, respectively. Note the XX photon shows a significantly better indistinguishability than the X photon. This could be due the fact that the X photon inherits an emission time uncertainty from the lifetime of the XX photon which is \SI{66.4}{ps}.

In summary, by pulsed two-photon resonant excitation of a quantum dot embedded in a CBG bullseye cavity, we have realized a deterministic entangled photon pair source with simultaneously an entanglement fidelity of 90$\%$, pair generation rate of 59$\%$, pair extraction efficiency of 62$\%$, and photon indistinguishability of 90$\%$ (93$\%$) without (with) temporal filtering. Future work is planned to apply electric field and surface passivation to reduce the blinking and spectral diffusion, which could improve both the source efficiency and photon indistinguishability over long time scale. Based on these, using active de-multiplexing with electro-optical modulators as demonstrated in multi-photon boson sampling \cite{Wang2017}, the entangled-photon pair source realized here can then be extended to multiple entanglement source. Future applications \cite{Kok2007,Browne2005} will include heralded multi-photon entanglement and boson sampling \cite{loss}, which can be performed without the complication of higher-order emissions, a notorious problem in parametric down-conversion \cite{Kwiat1995}. To improve the device yield, it is desirable to combine deterministic positioning for an optimal emitter-cavity coupling and\textit{ in-situ} strain tuning to minimize the fine structure splitting for the engineering of solid-state sources of photon pairs with near-unity degrees of entanglement, indistinguishability, and efficiency. Finally, our work can be extended to the realization of entanglement swapping \cite{swapping1993} between remote entangled photons from quantum dots embedded in CBG bullseye cavities, a step toward solid-state quantum repeaters \cite{Briegel1998}.

This work was supported by the National Natural Science Foundation of China, the Chinese Academy of Science, the Sci. Tech. Commission of Shanghai Municipality, the National Fundamental Research Program, the State of Bavaria. H.W., H.H., and T.-H.C. contributed equally to this work.

After the acceptance of this Letter, the XX and X single-photon count rate has been improved to $>$10 MHz. A preprint by J. Liu \textit{et al.} reporting CBG-coupled droplet GaAs quantum dots emitting at ~770 nm appeared on arXiv.1903.01339.



\subsection{$\longleftarrow$Supplemental Materials$\longrightarrow$}

\subsection{Characterization of the circular Bragg grating}

\begin{figure}[ht]
  \centering
  \includegraphics[width=0.485\textwidth]{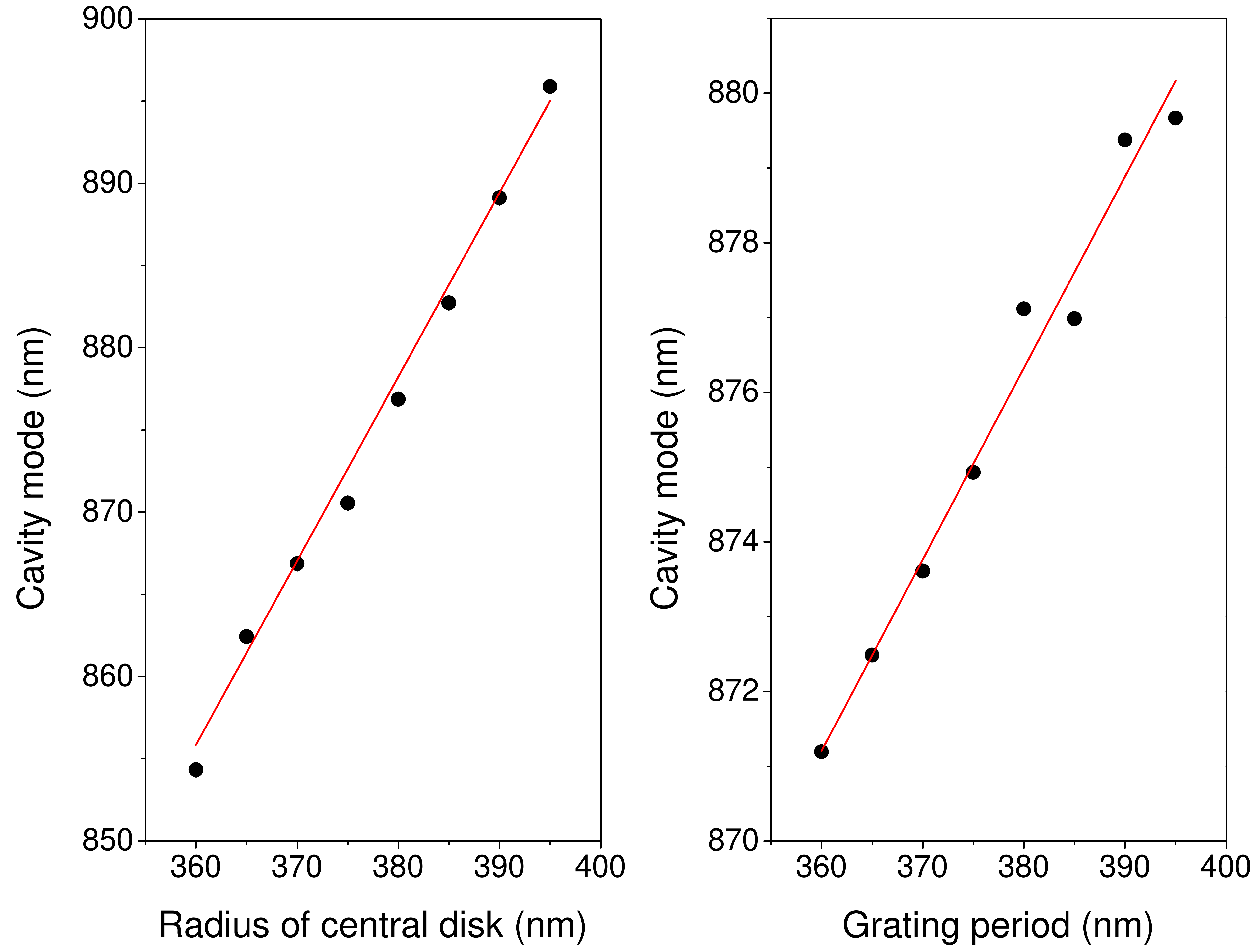}
  \caption{(a) The cavity mode shift as a function of the radius of central disk. (b) The cavity mode shift as a function of the period of surrounding grating.}
  \label{Fig:5}
\end{figure}

To determine an appropriate range of the cavity parameters, we performed two groups of testing experiments with CBG cavities setting at varying parameters. The most relevant parameters for the CBG cavity are: central disk, surrounding grating and trenches. We fix the width of trench to \SI{100}{nm}. The first group is that the period of the grating is fixed to \SI{365}{nm}, while the radii of the central disks are varied from \SI{360}{nm} to \SI{395}{nm} with a step of \SI{5}{nm}. The measured cavity modes are shown in Fig.~\ref{Fig:5}a, which shows that every \SI{1}{nm} change of the central disk radius will lead to a \SI{1.14}{nm} red shift of the cavity mode. Thus, precisely controlling the size of the central disk is essential to make a QD be resonant with cavity. In this experiment, we measured more than 10 CBG cavities with same parameter but in different position, and these data shows a derivation as small as \SI{0.87}{nm}, which is much smaller than linewidth of the cavity and indicates a good repeatability of the fabrication process. The second group consists of cavities with the fixed central disk radius of \SI{375}{nm} with the period of grating varying from \SI{360}{nm} to \SI{395}{nm} with a step of \SI{5}{nm}. As shown in Fig.~\ref{Fig:5}b, \SI{1}{nm} change of the grating period leads to a \SI{0.25}{nm} change of the cavity mode, which is much less sensitive than the case of the central disk.

\subsection{XX excited-state preparation efficiency and radiation efficiency}

Exploiting the intrinsic correlation in the XX-X cascaded two-photon emission, we put forward a way to directly and reliably estimate the XX excited-state preparation efficiency (the probability of the laser pulse actually pumps the quantum dot to the XX excited state) and quantum radiation efficiency (the probability the excited state decay actually results into electromagnetic radiation), which is the internal efficiency of the solid-state entangled-photon emitter. Without the need of additional measurements or assumptions, the estimation can be derived from the XX-X cross-correlation measurements such as that shown in Fig.~\ref{Fig:3} in the main text. The principle is similar to the absolute calibration of single-photon detector using correlated photon pairs from parametric down-conversion.

\begin{figure}
  \centering
  \includegraphics[width=0.465\textwidth]{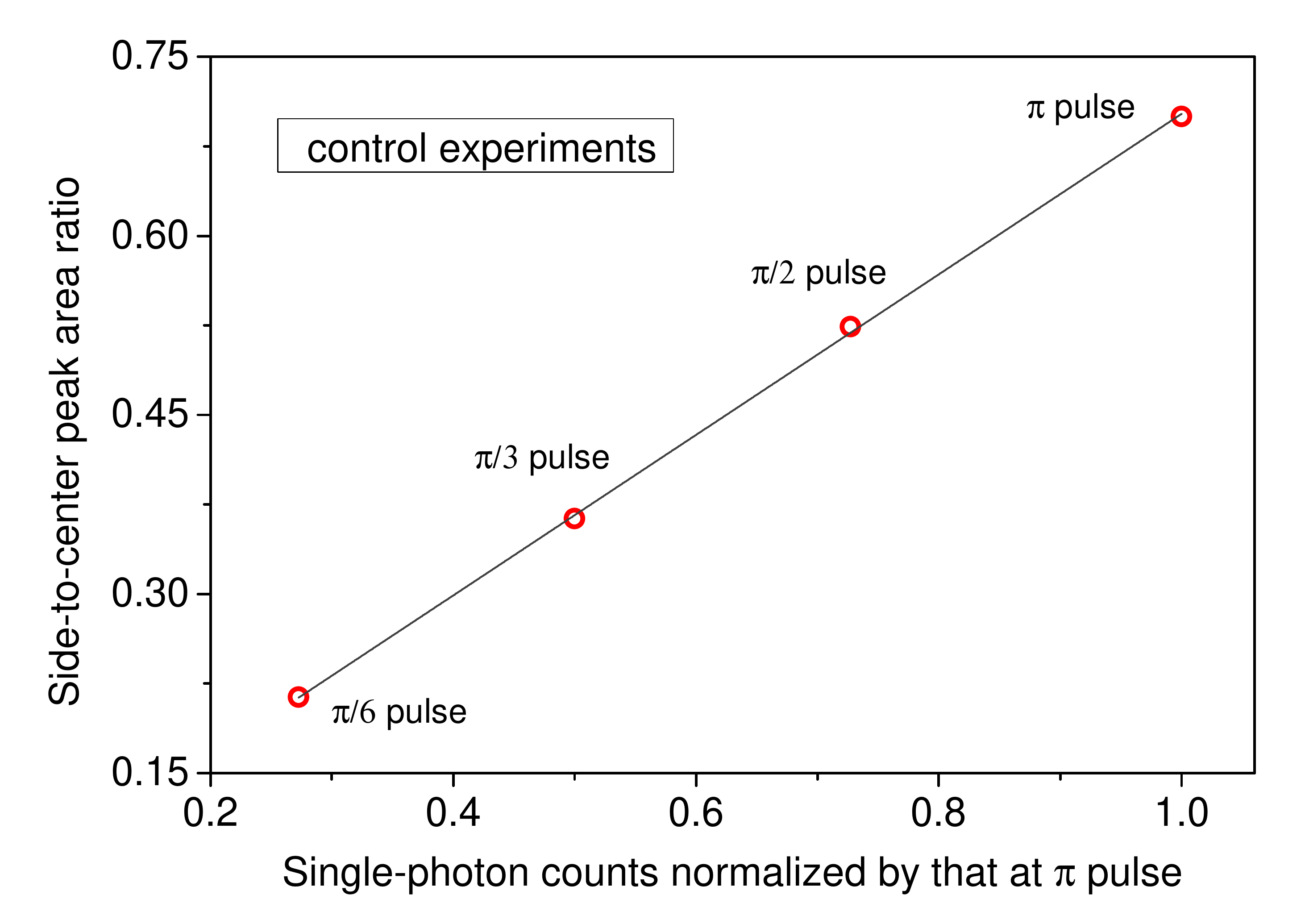}
  \caption{Absolute quantum efficiency calibration. Controlled experiment with optically tuned $p$, where the laser power is set to be at $\pi/6$, $\pi/3$, $\pi/2$ and $\pi$. In each setting, the side-to-center area ratio is extracted, which is perfectly fitted to a linear curve as a function of normalized single-photon counts.}
  \label{Fig:7}
\end{figure}

We assume the overall efficiency of the XX excited-state preparation and radiation probability is $p$. At zero time delay, the peak area is proportional to $p\eta_1\eta_2$, where $\eta_1$ and $\eta_2$ is the overall two optical path and detection efficiencies in the set-up. Here $p$ appears only once because the two cascaded emitted photons are correlated. However, the peak area at non-zero delay is proportional to $p^2\eta_1\eta_2$, because it is associated with the coincidence of two independent pulse excitation at different time (the two photons are not cascaded). We denote $A_{C}$ ($A_{S}$) as the sum of the peak areas at zero (non-zero) delay. Thus the XX excited-state preparation and radiation efficiency can be calculated as $p=A_{S}/A_{C}$, which is the side-to-center ratio. In our work, at $\pi$ pulse, the side-to-center ratio is $p=0.70(2)$.

To validate our method, we perform control experiments where we lower the excitation laser power from that corresponds to $\pi$ pulse to $\pi/2$, $\pi/3$, and $\pi/6$ pulses. The control of the laser power deliberately prepare the XX excited state with a relatively lower efficiency. We performed three others measurements at $\pi/2$, $\pi/3$, and $\pi/6$ pulses and extract the side-to-center peak area ratios which is plotted in Fig.~\ref{Fig:7}. A perfect linear dependence of the side-to-center peak area ratio to the single-photon counts at different laser power supports the correctness of our method.

\begin{figure}[h]
  \centering
  \includegraphics[width=0.43\textwidth]{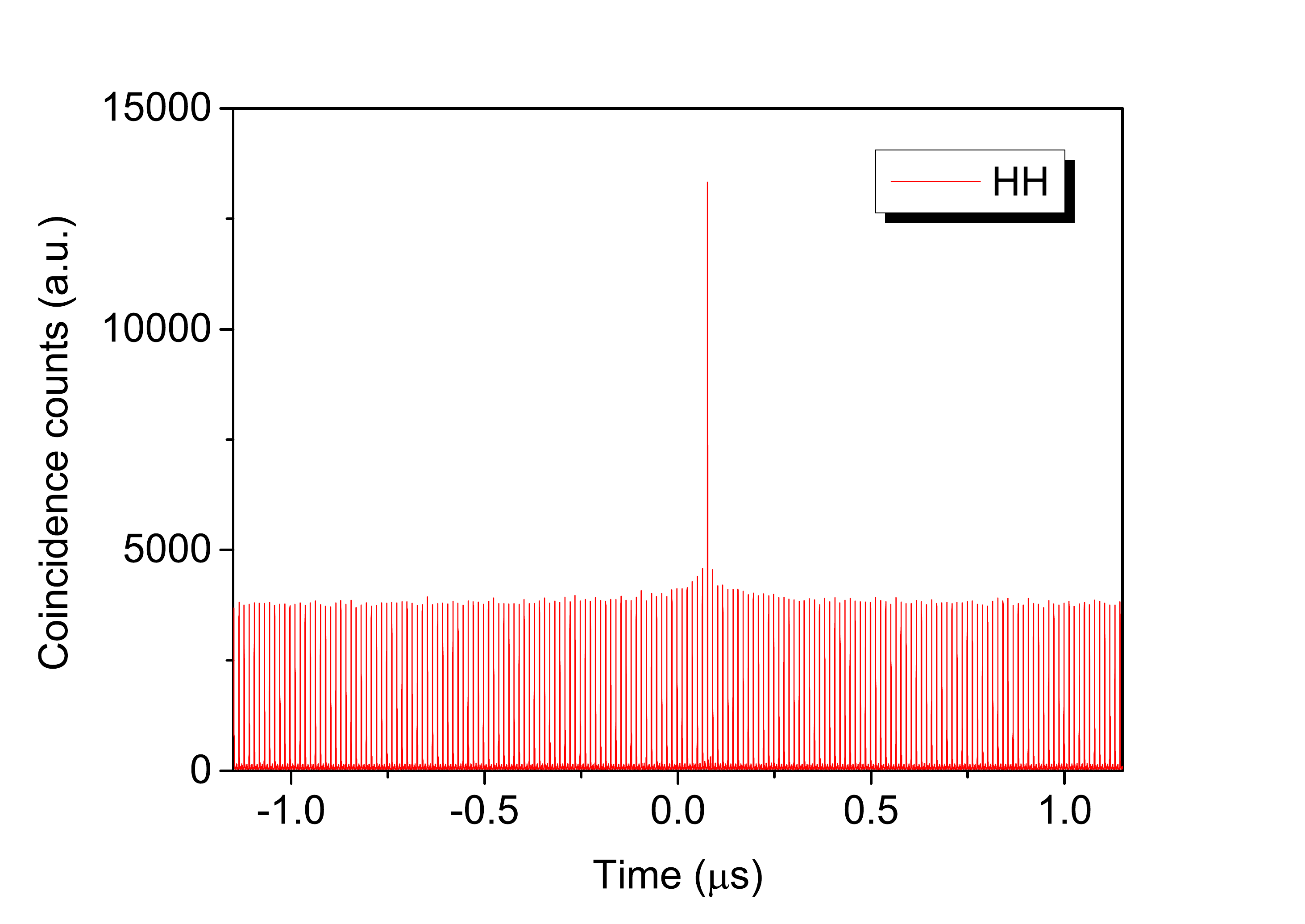}
  \caption{Long time scale cross-correlation of the entangled photons revealing a blinking of the investigated quantum dot.}
  \label{Fig:6}
\end{figure}

Figure~\ref{Fig:6} shows the coincidence decreases further at large time scale, indicating a blinking of the studied quantum dot, which might be caused by surrounding defects (the GaAs layer is thin, \SI{125}{nm}), and no passivation was applied. The blinking causes $\sim$16$\%$ loss.

\end{document}